\documentclass[twocolumn,astrosym,apj]{aastex631}
\usepackage{graphicx,amssymb,amsmath}    
\usepackage{amsthm,amsfonts,amscd,wasysym,cleveref}

\shorttitle{Little Red Dots: Red Cores within Blue Hosts}
\shortauthors{Ishikawa et al.}

\graphicspath{{./}{figures/}}
\newcommand{\progNum}{5664}

\newcommand{\targGNA}{GN-9771}
\newcommand{\targGNB}{GN-12839}
\newcommand{\targGNC}{GN-15498}
\newcommand{\targGND}{GN-16813}
\newcommand{\targGSA}{GS-13971}

\newcommand{\OIII}{[\textrm{O}~\textsc{iii}]}

\newcommand{\FeII}{\textrm{Fe}~\textsc{ii}}

\newcommand{\Lya}{${\rm Ly\alpha}$}
\newcommand{\Ha}{${\rm H\alpha}$}
\newcommand{\Hb}{${\rm H\beta}$}

\newcommand{\mum}{$\mu$m}

\newcommand{\zsys}{$z_{\textrm{sys}}$}

\begin{document}
\title{Spatial decomposition of Little Red Dots with JWST/NIRSpec IFU into broad-line red cores and narrow-line blue host galaxies}


\newcommand{\MKI}{MIT Kavli Institute for Astrophysics and Space Research, Massachusetts Institute of Technology, Cambridge, MA 02139, USA}
\newcommand{\MIT}{Department of Physics, Massachusetts Institute of Technology, Cambridge, MA 02139, USA}
\newcommand{\ISTA}{Institute of Science and Technology Austria (ISTA), Am Campus 1, 3400 Klosterneuburg, Austria}

\correspondingauthor{Yuzo Ishikawa}
\email{yishika2@mit.edu}

\author[0000-0001-7572-5231]{Yuzo Ishikawa}
\affiliation{\MKI}

\author[0000-0003-2895-6218]{Anna-Christina Eilers}
\affiliation{\MKI}
\affiliation{\MIT}

\author{Rohan P. Naidu}
\affiliation{\MKI}
\affiliation{Institute for Astronomy, University of Hawai‘i, 2680 Woodlawn Drive, Honolulu, HI 96822, USA}

\author[0000-0003-2871-127X]{Jorryt Matthee}
\affiliation{\ISTA}

\author[0000-0002-3120-7173]{Rongmon Bordoloi}
\affiliation{Department of Physics and Astronomy, North Carolina State University, Raleigh, NC 27695-8202, USA}

\author[0000-0002-0302-2577]{John Chisholm}
\affiliation{Department of Astronomy, The University of Texas at Austin, 2515 Speedway, Stop C1400, Austin, TX 78712,USA}
\affiliation{17 Cosmic Frontier Center, The University of Texas at Austin, Austin, TX 78712, USA}

\author{Jenny E. Greene}
\affiliation{Department of Astrophysical Sciences, Princeton University, 4 Ivy Lane, Princeton,NJ 08544, USA}

\author{Yilun Ma}
\affiliation{Department of Astrophysical Sciences, Princeton University, 4 Ivy Lane, Princeton,NJ 08544, USA}

\author[0000-0001-5851-6649]{Pascal A. Oesch}
\affiliation{Department of Astronomy, University of Geneva, Chemin Pegasi 51, 1290 Versoix, Switzerland}
\affiliation{Cosmic DAWN Center, Niels Bohr Institute, University of Copenhagen, Jagtvej 128, K\o benhavn N, DK-2200, Denmark}

\author{Wendy Q. Sun}
\affiliation{\MKI}
\affiliation{Institute for Astronomy, University of Hawai‘i, 2680 Woodlawn Drive, Honolulu, HI 96822, USA}`

\author[0000-0001-5586-6950]{Alberto Torralba}
\affiliation{\ISTA}

\author[0000-0003-1614-196X]{John R. Weaver}\thanks{Brinson Prize Fellow}
\affiliation{\MKI}

\author{Stijn Wuyts}
\affiliation{Department of Physics, University of Bath, Claverton Down, Bath BA2 7AY, UK}

\author[0000-0003-1207-5344]{Mengyuan Xiao}
\affiliation{Department of Astronomy, University of Geneva, Chemin Pegasi 51, 1290 Versoix, Switzerland}


\begin{abstract}
Little Red Dots (LRDs) are a population of compact red sources discovered by the James Webb Space Telescope (JWST). Imaging and spectroscopy have shown that LRDs exhibit a complex spectrum with a ``V-shaped'' continuum, broad Balmer emission lines, and in some cases Balmer absorption. While the physical origin of these components remains debated, recent studies propose that they arise from a compact central engine likely hosting a rapidly growing black hole embedded within a more extended host galaxy. We test this central engine + host galaxy model using JWST/NIRSpec integral field unit (IFU) spectroscopy to spectrally decompose the observed continuum, narrow and broad emission lines, and absorption. We spatially map each component for five broad \Ha-selected LRDs at $z\sim5$ observed with both the prism and high-resolution G395H grating. We find that the blue continuum emission is co-spatial with the narrow emission line region, while the red continuum arises from a compact core co-spatial with the broad Balmer emission and absorption. Spatial maps of the \OIII\ equivalent width reveal a pronounced decrease in the central core, consistent with this picture. Our work provides further evidence that the LRD emission is produced by at least two distinct physical components arising from a red central engine embedded within a blue host galaxy. 
\end{abstract}

\section{Introduction} \label{sec:intro}
Little Red Dots (LRDs) are a population of compact, red sources recently identified in JWST NIRCam surveys \citep[e.g.][]{Kocevski2023,Harikane2023,Kokorev2024,Greene2024,Williams2024,Matthee2024,Labbe2025,Akins2025,Kocevski2025}. These objects are often unresolved or only marginally resolved, suggesting compact sizes ($\leq 100-200\rm~pc$). They exhibit red, rest-frame optical colors with a detectable rest-ultraviolet (UV) component, producing a ``V-shaped'' continuum with a continuum break around the Balmer break at $\lambda_{\rm rest}\sim 3800\,\mathrm{\AA}$ \citep{Kocevski2023,Greene2024}. Spectroscopy revealed that these objects have both narrow emission lines (e.g.~\OIII, \Ha, \Hb) with $\sigma\sim100-300\rm~km~s^{-1}$ and broad Balmer emission lines with typical full-width-half-max (FWHM) of $1000-3000\rm~km~s^{-1}$ \citep[e.g.][]{Kocevski2023,Greene2024, Matthee2024,Furtak2023}. Interestingly, nearly 60\% of LRDs show Balmer absorption lines \citep{Matthee2024,Juodzbalis2024, Matthee2026,LinX2026a,LinX2026b}. Finally, these objects appear to be highly abundant with a number density around $10^{-5}\rm~cMpc^{-3}$  at $z\gtrsim4$ that then dramatically drops at $z<4$ \citep{Kocevski2023,Greene2024,Matthee2024,MaY2025,MaY2026,Park2026}.

Ever since their discovery, the nature of LRDs has been heavily debated. Early interpretations suggested that LRDs may represent massive, evolved stellar populations, supported by their blue rest-UV continua, Balmer absorption, and strong nebular emission, or accreting, dust-obscured massive black holes (active galactic nuclei; AGN), motivated by the presence of broad Balmer emission lines and narrow forbidden lines \citep[e.g.][]{Kocevski2023,Greene2024, Matthee2024,Furtak2023,Torralba2026b}. Some studies have instead proposed that LRDs may consist of composite systems, such as a blue star-forming host galaxy and a compact red AGN-like component \citep[e.g.][]{WangB2024,JiX2025,MaY2025,SunW2026}. However, no single scenario has yet reproduced the full set of observed properties, in particular the characteristic continuum shape \citep[e.g.][]{LeungG2025} together with the broad emission lines. Moreover, these interpretations are challenged by several conflicting observations, including the weak X-ray emission \citep{YueM2024a,Hviding2026}, the lack of far-infrared emission associated with a hot dusty torus \citep[e.g.][]{Akins2025,PerezGonzalez2024,WangB2024, XiaoM2025}, and the lack of variability over timescales expected for AGNs of comparable luminosity \citep{TeeWL2025,Burke2025,LiuZ2026}, although see \cite{Lambrides2026}.

These debates have spurred the development of alternative models. In particular, recent phenomenological models interpret LRDs as compact, AGN-like central engines embedded in a dense, optically thick gas envelope \citep[e.g.][]{Inayoshi2025a,Rusakov2026} residing in host galaxies. In these models, the gas envelope produces the red blackbody-like continuum with a characteristic temperature of around $T\sim5000\rm~K$ \citep{Naidu2025, deGraaff2025a, deGraaff2025b} and the Balmer emission and absorption \citep{Matthee2026}, while the UV continuum arises from a combination of emission from the central engine and the host galaxy (e.g.~Black Hole Star; \citealt{Naidu2025,SunW2026,Naidu2026}). Alternatively, \cite{Martins2026} show that both continuum and emission line features may arise self-consistently from a single dense atmosphere structure. A similar theoretical model interprets LRDs as quasi-stars, an accreting core-collapse black hole within a stellar photosphere \citep{Begelman2026}. Other models have proposed that LRDs may represent an early stage of massive black hole growth due to a direct collapse black hole \citep[e.g.][]{Pacucci2023, Inayoshi2025b} with possibly extreme super-Eddington accretion \citep[e.g.][]{Lambrides2024,Pacucci2024b,LiuH2025}. The variety of proposed models reflects the current uncertainty surrounding the physical origin of the observed continuum and emission line features.

Observational efforts have begun to probe the contribution of the host galaxy to the LRD emission. NIRCam imaging have revealed extended, blue, UV components in some LRDs, possibly tracing the host galaxy \citep{Killi2024, ZhangY2025, Yanagisawa2026,Cloonan2026} or UV-bright companions \citep{Baggen2025,Baggen2026}. VLT/MUSE observations have also revealed an extended \Lya\ halo around an LRD \citep{Torralba2026a}. However, not all LRDs show a detectable extended UV components \citep{Merida2025, Rinaldi2025a,Cloonan2026}. Spectral analyses further suggest that the diversity in the LRD spectra may be due to the varying contributions from the host galaxy and the central engine \citep{SunW2026,Barro2026}.

If LRDs are composed of a compact central engine and a host galaxy, their emission should exhibit distinct spatial and spectral signatures: we would expect the host galaxy to trace more extended structures, while the central engine should remain compact. A simultaneous spatial and spectral decomposition is therefore needed to directly map the origin of the continuum and emission lines. To investigate this, we use JWST Near-Infrared Spectrograph (NIRSpec; \citealt{Jakobsen2022}) IFU \citep{Boker2022} to simultaneously decompose the observed LRD emission, both spatially and spectrally. 

The outline of the paper is as follows. In Section~\ref{sec:dataRedux} we briefly describe our target selection, the observational setup, and data reduction. In Section~\ref{sec:specAnaly} we describe the LRD decomposition model (central engine + host) and the spatially-resolved spectroscopic analysis. Then, we present the spectral fit results and spatial maps in Section~\ref{sec:results}. We present the interpretations in Section~\ref{sec:discuss} and summarize our results in Section~\ref{sec:concl}. All spectral fits are performed with wavelengths in the vacuum scale. We adopt the $\Lambda$CDM cosmology with $h = 0.7$, $\Omega_M = 0.3$, and $\Omega_{\Lambda} = 0.7$. 

\begin{table*}
\centering
\caption{Observation summary of JWST/NIRSpec IFU. We estimate \zsys\ using the high-resolution \OIII\ emission lines and show the measurement uncertainties.} 
\label{tab:obsv} 
\begin{tabular}{ccccc}
    \hline
    Target  & RA & DEC & $z_{sys,\rm[O~\textsc{iii}]}$  \\
    -  & (J2000) & (J2000) & -  \\
    \hline
    GN-9771   & 12:37:07.44 & $+$62:14:50.31 & $5.53450 \pm 0.00008$  \\
    GN-12839  & 12:37:22.63 & $+$62:15:48.11 & $5.24114 \pm 0.00002$  \\
    GN-15498  & 12:37:08.53 & $+$62:16:50.82 & $5.08457 \pm 0.00007$  \\
    GN-16813  & 12:36:43.03 & $+$62:17:33.12 & $5.35853 \pm 0.00001$  \\
    GS-13971  & 03:32:33.20 & $-$27:47:24.90 & $5.48159 \pm 0.00002$  \\
    \hline
\end{tabular}
\end{table*}

\section{Observations and data reduction}\label{sec:dataRedux}
\subsection{Sample Selection}
We obtained NIRSpec IFU observations of five broad-line LRDs at $z>5$ as part of the JWST GO Cycle 3 program \#\progNum\ (PI: Matthee). Initial results from the program were presented in \cite{Torralba2026b} and \cite{Matthee2026}. The targets, listed in Table~\ref{tab:obsv}, were selected from the faint, broad \Ha-emitting LRD candidates identified by \citet{Matthee2024} in the NIRCam grism survey with no color pre-selection applied. The observed objects span a range of \Ha\ profiles, rest-UV and optical colors, and estimated SMBH masses. The IFU observations were completed with both the high-resolution G395H grating, which provides simultaneous coverage of \Hb\ and \Ha\ emission, and the low-resolution PRISM, which provide a broad wavelength coverage from the rest-UV to rest-optical. 

\subsection{Observation and Data Reduction}    

NIRSpec IFU observations were set up with the PRISM/CLEAR and G395H/F290LP. This results in an effective observed wavelength coverage of $2.87-5.27$ \mum\ with a wavelength-dependent spectral resolution between $R=30-300$ (PRISM) and $R=2700-3500$ (G395H). There were no dedicated target verification exposures. We used the NRSIRS (PRISM) and NRSIRS2RAPID (G395H) readout modes and a medium cyclic 8-point dither pattern to improve the spatial sampling of the point-spread-function (PSF). We summarize the targets observed in Table~\ref{tab:obsv}. 

The IFU data reduction roughly follows the treatment in \cite{Vayner2023} and \cite{Ishikawa2025a}. The data reduction was completed using the STScI JWST pipeline\footnote{\url{https://github.com/spacetelescope/jwst}} version 1.17.1 with Calibration Reference Data System version 12.0.9 (\texttt{jwst\_1299.pmap}). The first stage, \texttt{Detector1Pipeline}, performs standard infrared detector reductions. We correct for the $1/f$ noise \citep{Schlawin2020} on each detector (NRS1 and NRS2, where appropriate) by using a running mean algorithm.  We also use \texttt{snowblind}\footnote{\url{https://github.com/mpi-astronomy/snowblind}} to remove noise from snowball effects and cosmic rays. Then, these rate files are processed by \texttt{Spec2Pipeline} that produces calibrated spectra data, assigns the world coordinate system, and extracts the 2D spectra to build a 3D datacube using \texttt{drizzle}. Finally, we apply a sigma clip routine to mask pixels with extreme outliers and use the Photutils \texttt{reproject} method \citep{Vayner2023} to align and combine the different dither exposures into a single datacube with a spatial resolution of $0.05\arcsec$ per spaxel. 

Since no dedicated background exposures were taken, we perform aperture background subtraction from the datacubes. Both the PRISM/CLEAR and G395H/F290LP datacubes showed spatially and spectrally varying background. We estimated the spatially varying background in each wavelength slice using the Photutils \texttt{Background2D} routine. Sources were masked using circular apertures, and the background was computed in $3\times3$ pixel tiles using a sigma-clipped median estimator. The resulting background maps were smoothed with a $3\times3$ box filter and subtracted from each slice before reconstructing the final background-subtracted datacube. Lastly, we mask out all foreground sources with circular apertures from the fully calibrated datacubes before the subsequent spectral analyses. 

\begin{figure}
    \begin{center} 
    \includegraphics[width=\columnwidth]{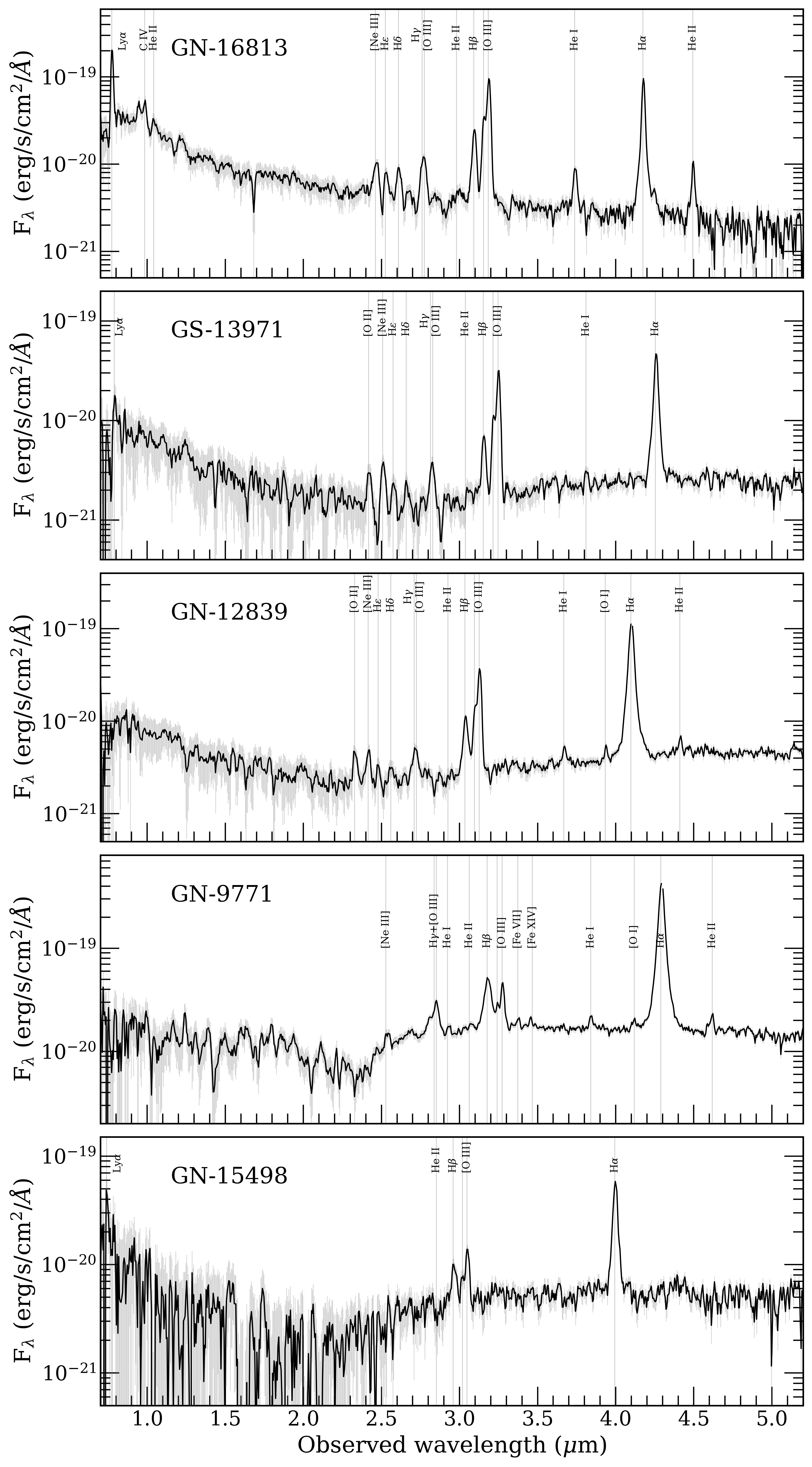}
    \end{center}
    \caption{Aperture spectra, extracted within $r=0.1''$ with the PRISM mode, centered on each LRD reveals a diversity in their continuum shapes. Most notably, the characteristic V-shape becomes progressively less pronounced as the prominence of the rest-UV component increases. We show the sample spectra in order of UV prominence and notable emission lines.}
    \label{fig:prism}
\end{figure}

\begin{figure*}
    \begin{center} 
    \includegraphics[width=0.95\textwidth]{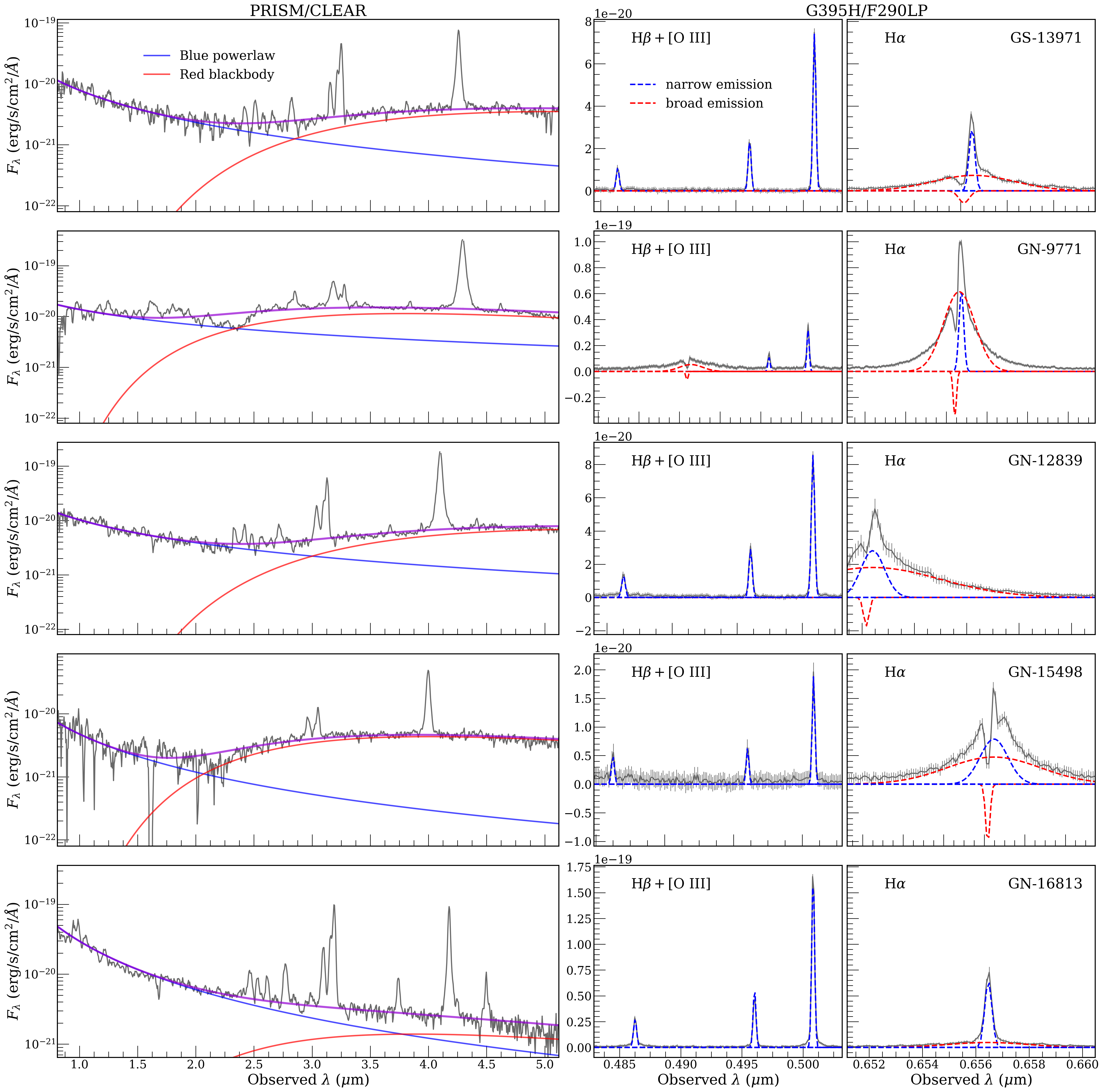}
    \end{center}
    \caption{We show sample fits of the PRISM continuum (left) and G395H spectral line fits around \Hb-\OIII\ (center) and \Ha\ (right) for each target. The PRISM continuum is fit with a two-component model: blue powerlaw and red blackbody. The narrow lines are shown in blue, and the broad lines and absorption are shown in red. Since the \Ha\ profile of \targGNB\ initially fell into the NIRSpec chip gap, the target was offset from the IFU center to partially recover the emission line, resulting in a truncated line profile. The targets are listed in order of RA.}
    \label{fig:spectfits}
\end{figure*}

\section{Spectral Analysis}\label{sec:specAnaly}
The goal of this work is to understand the spatial origin of each spectral components of LRDs. To this end, we decompose each LRD spectrum into different spectral components: a blue continuum that largely dominates the rest-UV until the Balmer Break at $\lambda_\mathrm{rest}=0.4\mathrm{\mu m}$, a red continuum that dominates redward of the Balmer Break, the narrow emission lines (\Ha, \Hb, and the \OIII$\lambda\lambda 4959,5007$ doublet), broad Balmer emission lines (\Ha\ and \Hb), and the associated Balmer absorption. We treat and fit each emission lines (narrow and broad) as kinematically and spatially independent components. 

Our objective is to construct spatially-resolved maps of the continuum (intensity) and spectral lines (intensity, velocity offset $\Delta v$, and velocity dispersion $\sigma_v$). Spectral line fits are performed on a spaxel-by-spaxel basis in the velocity space relative to the systemic redshift \zsys. 
The \zsys\ is determined by fitting the observed \OIII\ doublet in the spectrum extracted
with an $r=0.1''$ aperture, centered on the continuum core, from the G395H datacubes, and is fixed for each target, summarized in Table \ref{tab:obsv}.  Each emission and absorption line is modeled as a Gaussian profile parameterized by the peak flux density $F_{\rm pk}$ (in $\rm erg~s^{-1}~cm^{-2}~\AA^{-1}$), the line-of-sight velocity offset $\Delta v$ (in $\rm km~s^{-1}$), and the velocity dispersion $\sigma_v$ (in $\rm km~s^{-1}$). All fits are performed with \texttt{LMFIT} (version 1.3.2; \citealt{lmfit2024}). 

\begin{figure*}
    \begin{center} 
    \includegraphics[width=0.85\textwidth]{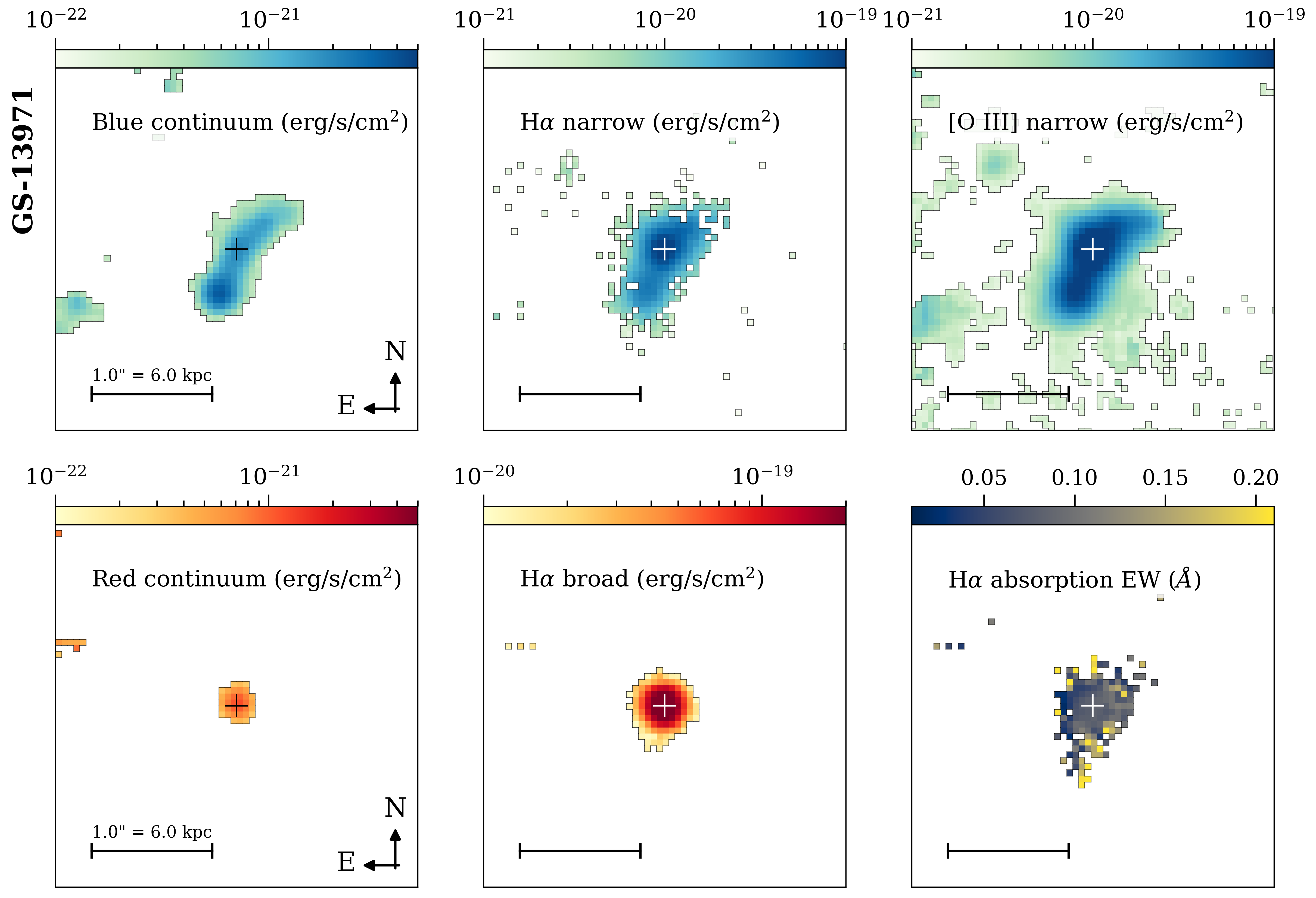}
    \end{center}
    \caption{We show the integrated intensity maps of the blue continuum (top left), narrow \Ha\ emission (top center), the narrow \OIII\ emission (top right), red continuum (bottom left), and the broad \Ha\ emission (bottom center) for \targGSA. We also show the $\textrm{EW}_{H\alpha,\rm abs}$ map of the \Ha\ absorption (bottom right). All maps are constructed from spaxels with $\mathrm{SNR}>3$. The cross indicates the centroid location of the LRD. }
    \label{fig:gs13971}
\end{figure*}

\subsection{Decomposing the blue and red continuum}
Aperture spectra of the LRDs, shown in Figure~\ref{fig:prism} reveal a diversity in their continuum shapes. A notable trend is the varying contributions from the blue and red continuum components, resulting in differences in the prominence of the V-shape. We model the blue and red part of the continuum emission in two different ways. 
First, we treat the blue component as a powerlaw, while the red continuum is described as a red modified blackbody  \citep[e.g.][]{deGraaff2025a}. This model is defined as follows:
\begin{equation}
    F_{\rm PRISM,\lambda} = A_{\textrm{PL}} \bigg(\frac{\lambda}{\lambda_A} \bigg)^{\alpha} + A_{\textrm{BB}} B_{\lambda}(\lambda,T) \bigg(\frac{\lambda_B}{\lambda}\bigg)^{\beta}.
\end{equation}
The powerlaw term is defined by the flux normalization $A_{\textrm{PL}}$ at rest-frame wavelength $\lambda_A=1500\,\mathrm{\AA}$ 
and the spectral index $\alpha$. The blackbody term is defined by the Planck function $B_{\lambda}(\lambda,T)$ at a temperature $T$, modified by a powerlaw slope $\beta$ and an amplitude $A_{\textrm{BB}}$ at $\lambda_B=5500\,\mathrm{\AA}$. We set a prior on $\alpha$ between $(0,-5)$ and on $T$ between $\rm(10^3~K,10^5~K)$.

We also decompose the continuum based on the empirically-derived LRD models, in which the LRD continuum is composed of a blue star-forming galaxy and the ``Black Hole Star'' central engine,  detailed in \cite{SunW2026}. The key difference between the two models is that the latter takes fixed spectral shapes that is scaled to the fit the data at each spaxel, while the former model parameters also include the spectral slope and blackbody temperature. 

We fit the observed PRISM spectrum at each spaxel in the IFU datacube with the described model. First, we mask out major UV and optical emission lines, including the \FeII\ forest. Then we fit the spectrum at each spaxel with both models, fitting for their respective components. In the left panels of Figure~\ref{fig:spectfits} we show example spectral decompositions of the blue/red continuum taken from select spaxels of each source. Once we determine the best-fit continuum models, we integrate each of the blue and red continuum to produce broadband intensity maps. We find that both continuum models produce similar fits and maps. Henceforth, we present results based on the first model (powerlaw+blackbody). We show the decomposed continuum maps in the left panels of Figure~\ref{fig:gs13971} for \targGSA\ and the remainder of the sample in Appendix~\ref{apdx:Map}.

\subsection{Decomposing the emission and absorption lines}

With the high-resolution G395H mode, we can resolve the key optical emission lines. We define the spectral model as a sum of the underlying continuum ($F_{\rm cont}$), narrow-lines (\Ha, \Hb\, and \OIII; $F_{\rm i,nl}$), broad-lines (\Ha\ and \Hb; $F_{\rm bl}$), and the \Ha\ absorption ($F_{\rm abs}$): 
\begin{equation}
    F_{\rm G395H,\lambda} = F_{\rm cont}+\Sigma F_{i, \rm nl}+\Sigma F_{i,\rm bl}+F_{i, \rm abs}
\end{equation}

We fit for $F_{\rm G395H,\lambda}$ at each spaxel and generate integrated line intensity and kinematic ($\Delta v$ and $\sigma_v$) maps of each spectral component. We also compute the equivalent width of the \Ha\ absorption ($\textrm{EW}_{H\alpha,\rm abs}$) by setting the baseline flux as the sum of the continuum, narrow-line \Ha, and broad-line \Ha\ models:
\begin{equation}
    \textrm{EW}_{H\alpha,\rm abs} = \int\Big(1-\frac{F_{\rm\lambda}}{F_{\rm cont}+F_{\rm H\alpha,nl}+F_{\rm H\alpha,bl}}\Big)d\lambda.
\end{equation}

The \Ha\ for \targGNB\ unfortunately landed on the edge of the detector gap, resulting in a partial truncation of the broad \Ha\ component. To stabilize the fit, we tied the velocity offsets of the narrow and broad \Ha\ lines, while the absorption component is allowed to vary freely with respect to the emission lines. 

\begin{figure*}
    \begin{center} 
    \includegraphics[width=\textwidth]{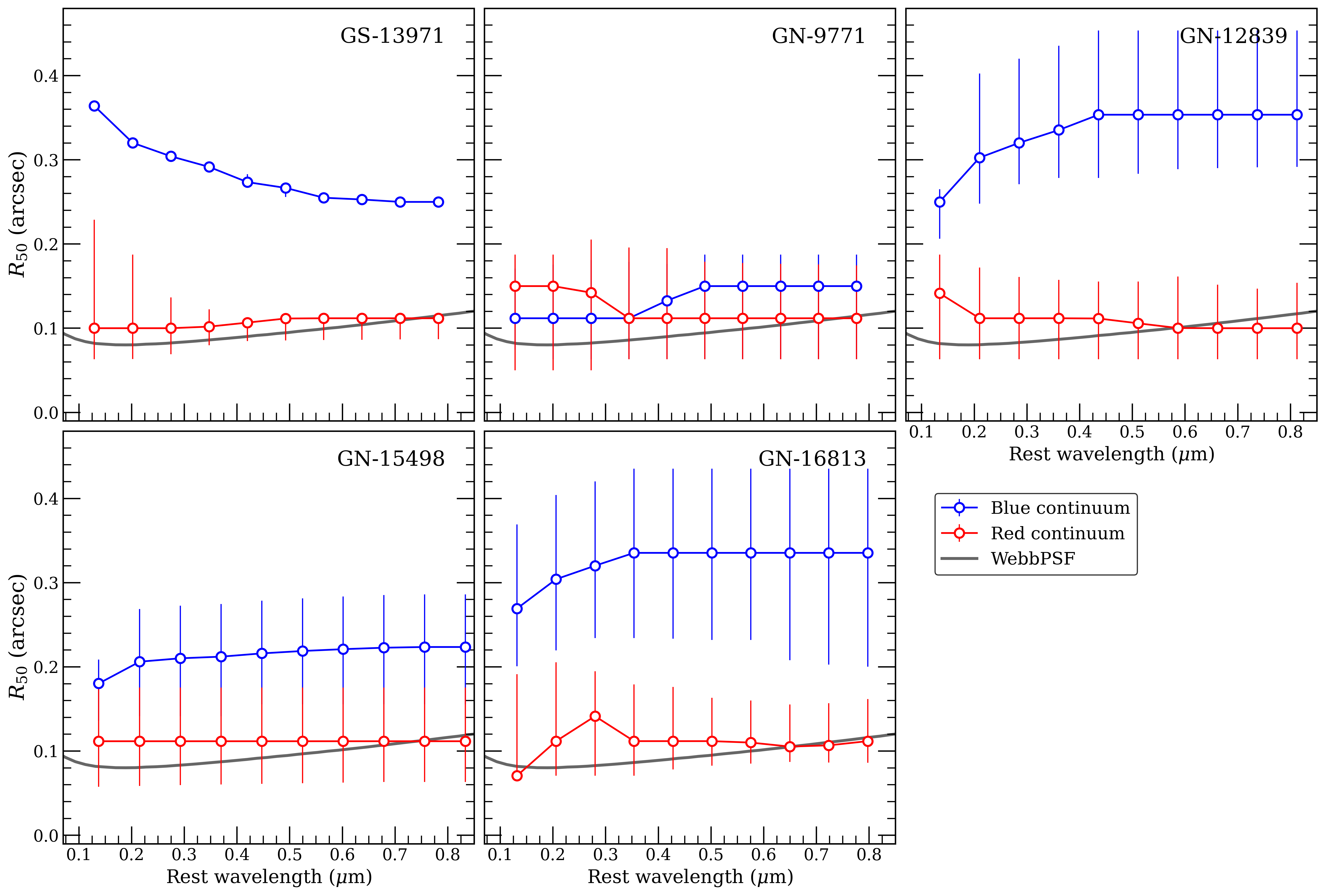}
    \end{center}
    \caption{We show the wavelength-dependent $R_{1/2}$ measurements of the blue continuum and the red continuum, measured within $\Delta\lambda\approx0.75~\mathrm{\mu m}$ bins. We find that the red component is unresolved and the blue component is extended. We also compare the convolved theoretical PSF at each wavelength bin, measured with \texttt{STPSF} (WebbPSF).}
    \label{fig:waveSBplot}
\end{figure*}

\begin{figure*}
    \begin{center} 
    \includegraphics[width=0.97\textwidth]{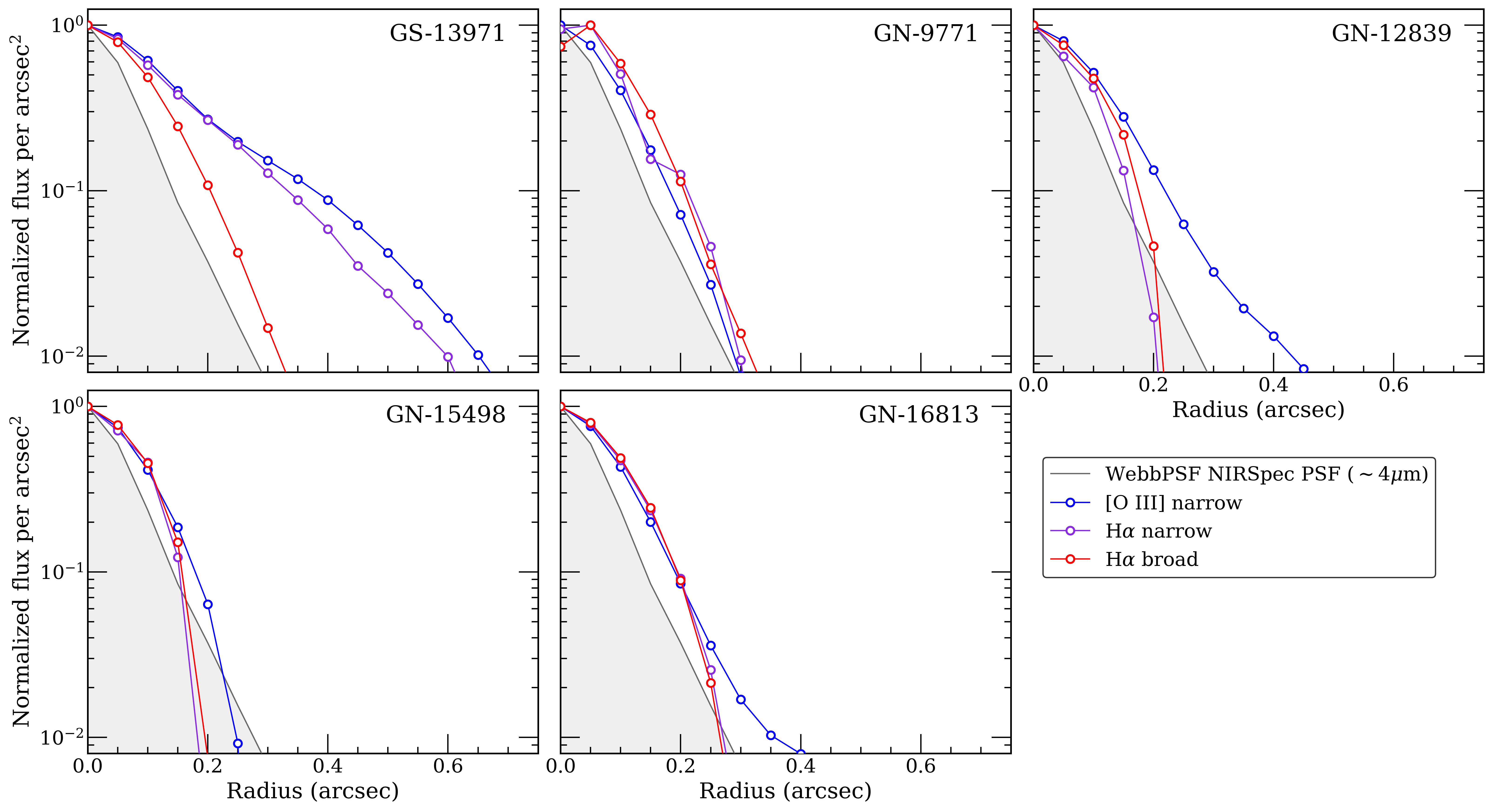}
    \end{center}
    \caption{We show normalized radial surface brightness profiles of the narrow-line emission (\Ha\ and \OIII\ in blue/purple) and the broad-line \Ha\ emission (red). We compare the intensity profiles with the theoretical JWST PSF profiles shown in shaded gray. We find that the broad \Ha\ is consistently compact, whereas the \OIII\ shows both compact and extended sizes.}
    \label{fig:SBplot}
\end{figure*}

\begin{figure*}
    \begin{center} 
    \includegraphics[width=0.95\textwidth]{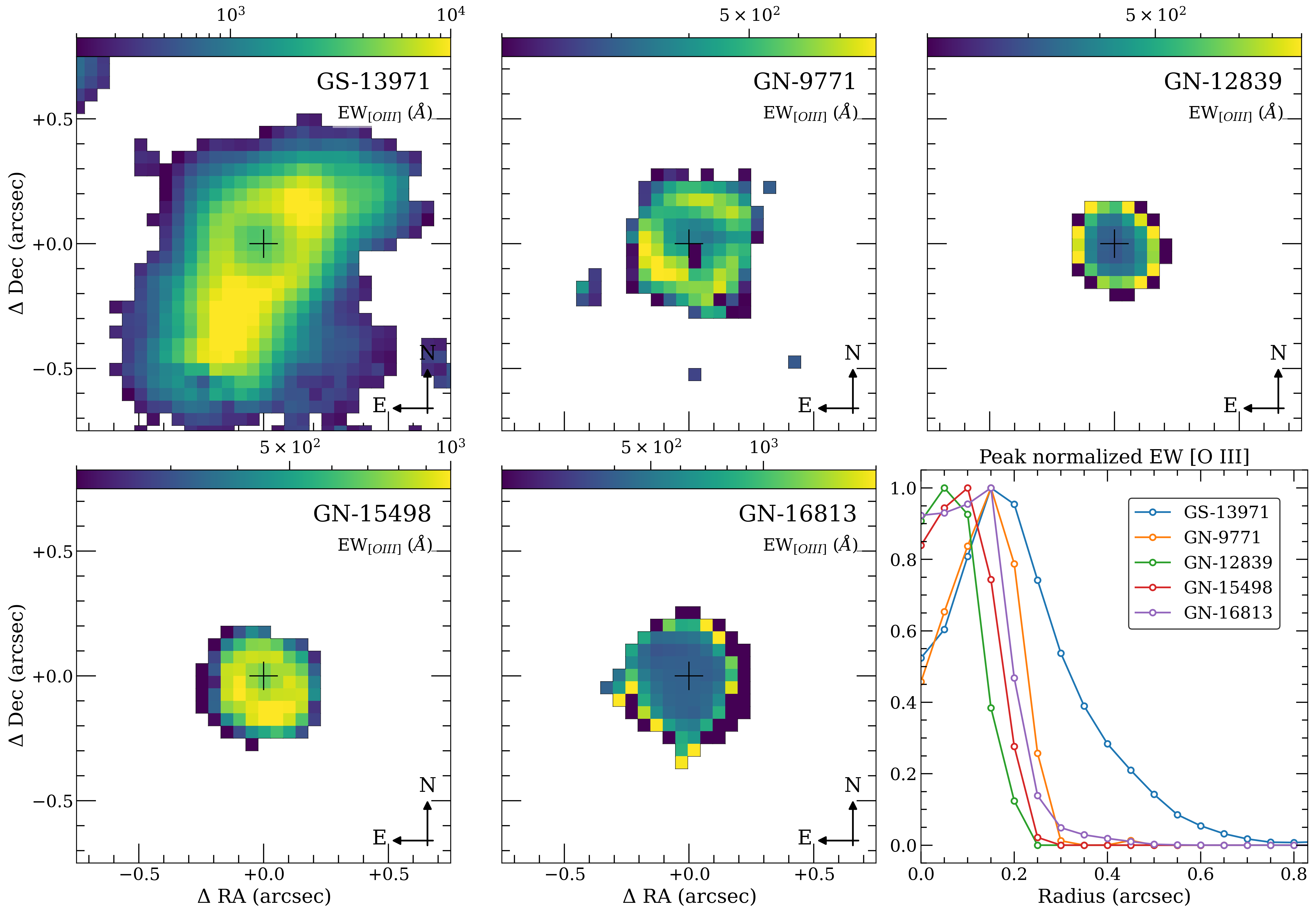}
    \end{center}
    \caption{We show \OIII\ equivalent width maps for all targets, zoomed in so that each box represents $1.5''\times1.5''$. All targets show a non-monotonic radial profile with a central dip or plateau in $\textrm{EW}_{\OIII}$, where the underlying red continuum is stronger than the \OIII\ emission, followed by a slight increase in \OIII\ emission, before radially decreasing outward shown in the radial $\textrm{EW}_{\OIII}$ profile (bottom right), normalized for comparison.}
    \label{fig:EWmaps}
\end{figure*}

\section{Results}\label{sec:results}
\subsection{Decomposed continuum maps}\label{sec:rescont}
The continuum decompositions reveals spatially extended blue continuum and compact red continuum components. Visually, \targGSA, \targGNB, and \targGND\ show more extended blue component, whereas the distinction is less clear for \targGNA\ and \targGNC. We quantify the spatial extent of the blue and red continuum by measuring their non-parametric half-light radii, $R_{1/2}$, defined as the radius enclosing 50\% of the total flux, as a function of wavelength. We divide the continuum model fits into wavelength bins with size $\Delta\lambda\approx0.75~\mathrm{\mu m}$, collapse the model into a single image at each bin, and measure $R_{1/2}$ using a fixed-aperture curve-of-growth analysis. Uncertainties are defined as the 16th to 84th percentile range of the resulting $R_{1/2}$. We plot the $R_{1/2}$ vs.~rest-$\lambda$ relations of the blue and red continuum for each LRD, as shown in Figure \ref{fig:waveSBplot}. 

Although both the blue and red components are relatively compact, constrained within a radius of $0.5''$, several sources (\targGSA, \targGNB, \targGND, and to a lesser extent \targGNC) show systematically larger blue continuum sizes compared to the red continuum, while \targGNA\ shows comparable sizes due to the marginally resolved continuum. We note that some of the extended emission in \targGSA\ may originate from nearby companions (see Appendix~\ref{apdx:merger}).

\subsection{Decomposed emission line maps}\label{sec:resline}

In addition to the continuum decomposition, we have spatially-resolved decompositions of the ionized emission lines. We show select spectral decompositions of emission and absorption lines in the right panels of Figure~\ref{fig:spectfits}. We also show the spatial maps of the narrow \OIII\ and \Ha\ emission, the broad \Ha\ emission, and the \Ha\ absorption of \targGSA\ in Figure~\ref{fig:gs13971}; the remainder of the sample is shown in Appendix~\ref{apdx:Map}. We only show spaxels with $\mathrm{SNR}>3$. 

All five LRDs exhibit strong narrow-line \OIII\ and \Ha\ emission and broadened \Ha\ emission. Four targets show blueshifted \Ha\ absorption (\targGNA, \targGNB, \targGNC, and \targGSA), of which \targGNA\ and \targGSA, though noisier, also show \Hb\ absorption features (also see \citealt{Matthee2026}). No broad \OIII\ components are detected in any of the targets with typical velocity dispersions of $\sigma\sim80\rm~km~s^{-1}$. Of the five LRDs in the sample, \targGSA\ shows the clearest detection of extended ionized gas structures, with some contributions from nearby companions, discussed further in Appendix~\ref{apdx:merger}. 

Visual comparisons of the spectral component shows that the narrow \OIII\ emission and the broad \Ha\ emission have different spatial distributions. We quantify the spatial extent of the emission lines by plotting the radial surface brightness profiles of the emission lines in Figure~\ref{fig:SBplot}. We compare these radial profiles of the broad \Ha, narrow \Ha, and narrow \OIII\ with the theoretical JWST PSF models generated with \texttt{STPSF} \citep{Perrin2014}. First, we find that the broad \Ha\ is more compact and centrally concentrated with with a FWHM of $\sim0.19\pm0.05''$, roughly consistent with the JWST PSF ($\rm FWHM\sim0.13''$). Second, we see a variation in the morphology of \OIII: some show extended structure (\targGSA, \targGNB, and to a lesser extent \targGND) and others are compact, indistinguishable from the broad emission (\targGNA\ and \targGNC), with FWHM ranging between $0.18''$ to $0.25''$. What we see is a compact broad-line emitter that is consistent with a PSF, and an extended narrow-line emitter, best traced with \OIII.

\section{Discussion}\label{sec:discuss}

Based on the spatial and spectral decomposition, we propose that LRDs are composed of two key structures: a broad-line emitting, red continuum core (i.e.~the black hole central engine) and a narrow-line emitting, extended blue continuum (i.e.~the host galaxy). The connection is made by correlating the spatial extent of each component. Here we combine the results of the continuum and emission line decomposition to build a comprehensive picture of LRDs. %

The key question is whether the \OIII\ emission traces the blue continuum host galaxy. \cite{deGraaff2025b} and \cite{SunW2026} showed that the equivalent width of the \OIII\ emission ($\textrm{EW}_{\OIII}$) can serve as a proxy for the host fraction, in which larger $\textrm{EW}_{\OIII}$ values correspond to a greater host fraction. If this interpretation is correct, \OIII\ should approximately trace the host morphology and exhibit an anti-correlation with the red continuum source. To test this, we construct  $\textrm{EW}_{\OIII}$ maps from the spectral fits (Figure~\ref{fig:gs13971} and Appendix~\ref{apdx:Map}), as shown in Figure~\ref{fig:EWmaps}. Although the measured $\textrm{EW}_{\OIII}$ values vary for each source, we consistently see a non-monotonic radial profile, which shows a central dip in $\textrm{EW}_{\OIII}$ at $r<0.2''$ (relatively weaker \OIII\ compared to the red continuum) that increases with radii before decreasing radially outward. Visually, this results in an $\textrm{EW}_{\OIII}$-ring centered on the LRD. This behavior suggests that the inner regions are dominated by the red continuum core, while enhanced $\textrm{EW}_{\OIII}$ at larger radii indicates a spatially extended narrow-line-producing source. This spatial anti-correlation of the $\textrm{EW}_{\OIII}$ with the red core supports a scenario in which \OIII\ originates from a more extended source like a host galaxy, rather than the compact red source. 

We also find a correlation between the detection of the blue continuum and extended \OIII\ emission. This trend is clearly seen in \targGSA, \targGNB, and to a lesser extent \targGND. In contrast, \targGNA\ and \targGNC\ do not show a clear morphological separation of the blue/red continuum. Nevertheless, our continuum decompositions are broadly in agreement with \cite{Cloonan2026}, who showed that the LRD sizes depend on wavelength. Even so, the $\textrm{EW}_{\OIII}$ profiles in Figure~\ref{fig:EWmaps} suggest that the \OIII-emitting regions may still be marginally resolved in all five sources. This further supports the interpretation that the narrow-line emission is spatially associated with the blue host-galaxy component rather than the compact red continuum source. However, we note that some of the gas may still be photoionized by the central engine. Furthermore, the detectability and resolvability of the host galaxy likely depend on the contrast between the host and the central engine, making the host galaxy more difficult to detect and spatially resolve as the central engine becomes more dominant. Lastly, we note that some fraction of the blue continuum may itself originate from the central engine that may not be fully separated in our decomposition  \citep[e.g.][]{SunW2026}. 

\section{Summary}\label{sec:concl}
We have presented JWST/NIRSpec IFU observations of five LRDs using the NIRSpec IFU with the PRISM and G395H grating. We decompose the observed LRD spectra at each spaxel into different spectral components: a blue power-law continuum, a red modified blackbody continuum, broad emission lines, narrow emission lines, and the Balmer absorption. We then construct intensity maps of all components. 

As shown in Figures~\ref{fig:gs13971}, \ref{fig:waveSBplot}, and \ref{fig:SBplot}, we find that the red continuum and broad emission lines in all five targets are compact, while the blue continuum and narrow lines are generally more spatially extended, although the sizes of the blue and red components in two of the five objects are comparable. This implies that the size relation between the blue and red components is wavelength-dependent. Also, the \OIII\ equivalent width maps of the targets show a clear decrease in the center, suggesting that the \OIII\ does not trace the compact red continuum.

Overall, our results provide evidence that LRDs are composed of a broad-line emitting, red continuum core (e.g.~a central engine plausibly powered by a black hole) and a narrow-line emitting, extended blue continuum likely arising from the surrounding host galaxy, as recently suggested by \cite{Naidu2025}, \cite{SunW2026}, and others. 

\begin{acknowledgments}
This work is based on observations made with the NASA/ESA/CSA James Webb Space Telescope. The data were obtained from the Mikulski Archive for Space Telescopes at the Space Telescope Science Institute, which is operated by the Association of Universities for Research in Astronomy, Inc., under NASA contract NAS 5-03127 for JWST. These observations are associated with program \#\progNum. Some/all of the data presented in this article were obtained from the Mikulski Archive for Space Telescopes (MAST) at the Space Telescope Science Institute. The JWST data can be accessed via \dataset[DOI]{https://doi.org/10.17909/dfvx-tn24}.

Support for program \#\progNum\ was provided by NASA through a grant from the Space Telescope Science Institute, which is operated by the Association of Universities for Research in Astronomy, Inc., under NASA contract NAS 5-03127.

J.M. and A.T. acknowledge funding by the European Union (ERC, AGENTS, 101076224).

P.O. acknowledges that this work has received funding from the Swiss State Secretariat for Education, Research and Innovation (SERI) under contract number MB22.00072, as well as from the Swiss National Science Foundation (SNSF) through project grant 200020\_2073 49. The Cosmic Dawn Center (DAWN) is funded by the Danish National Research Foundation under grant DNRF140. 

J.R.W. acknowledges that support for this work was provided by The Brinson Foundation through a Brinson Prize Fellowship grant.

\end{acknowledgments}

%

\facilities{JWST}


\software{\texttt{astropy} \citep{astropy2013,astropy2018,astropy2022},
\texttt{LMFIT} \citep{lmfit2024},
\texttt{STPSF} \citep{Perrin2014},
\texttt{Photutils} \citep{photutils2024},
}

\bibliography{zBIB}{}
\bibliographystyle{aasjournal}

\appendix
\section{Intensity maps of other targets}\label{apdx:Map}
We show the spatial decomposition of the broad-line, narrow-line, absorption, and continuum components for \targGNA, \targGNB, \targGNC, and \targGND. Although the extent of the blue continuum and narrow-line emission vary, all targets consistently reveal a compact broad-line emitting, red continuum core, as shown in Figures~\ref{fig:gs13971}, \ref{fig:waveSBplot}, and \ref{fig:SBplot}. 

\begin{figure*}[h]
    \begin{center} 
    \includegraphics[width=0.8\textwidth]{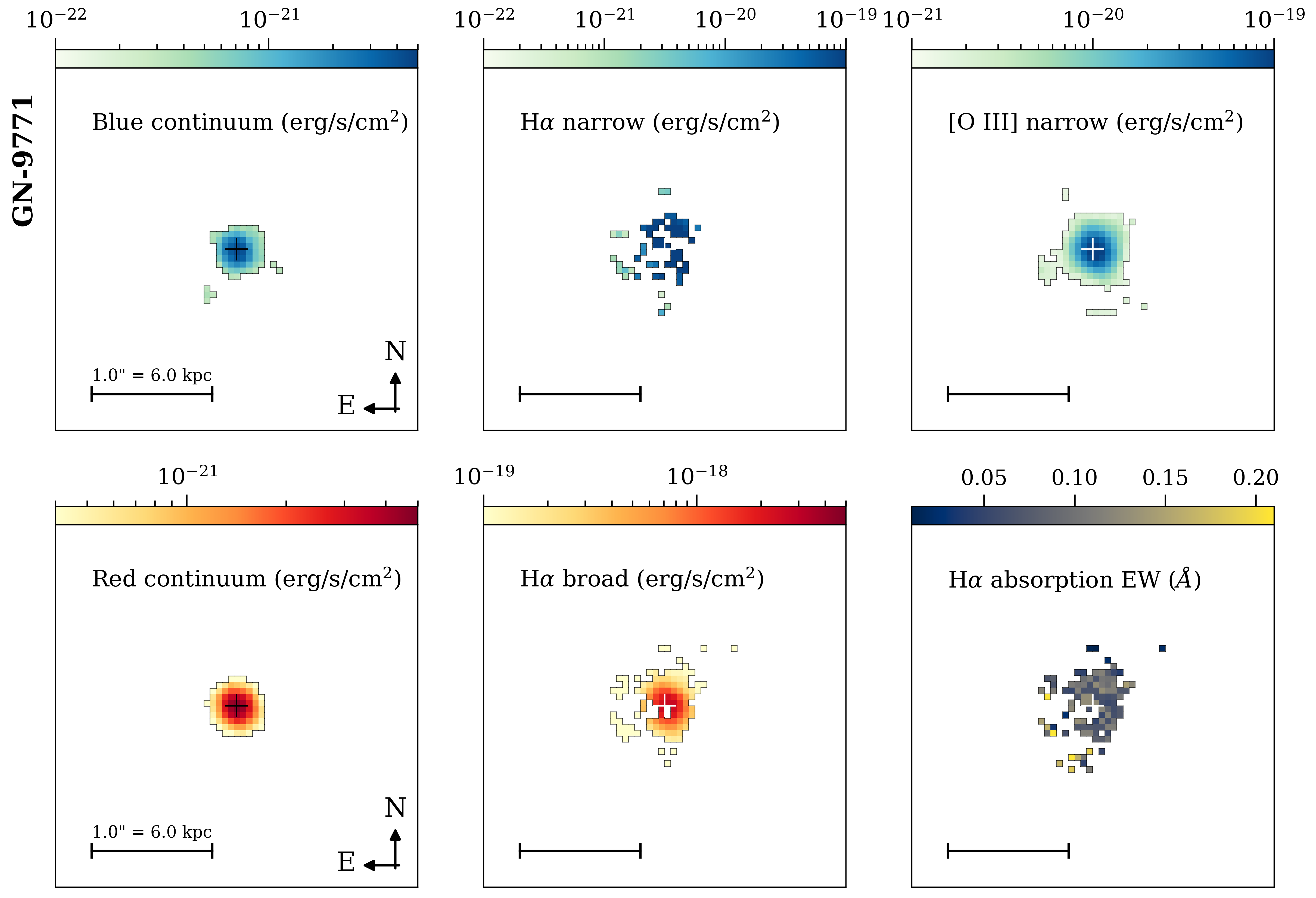}
    \end{center}
    \caption{We show the spatial intensity maps of \targGNA. The panel layout and descriptions are identical to those in Figure~\ref{fig:gs13971}.}
    \label{fig:gn9771}
\end{figure*}

\begin{figure*}
    \begin{center} 
    \includegraphics[width=0.8\textwidth]{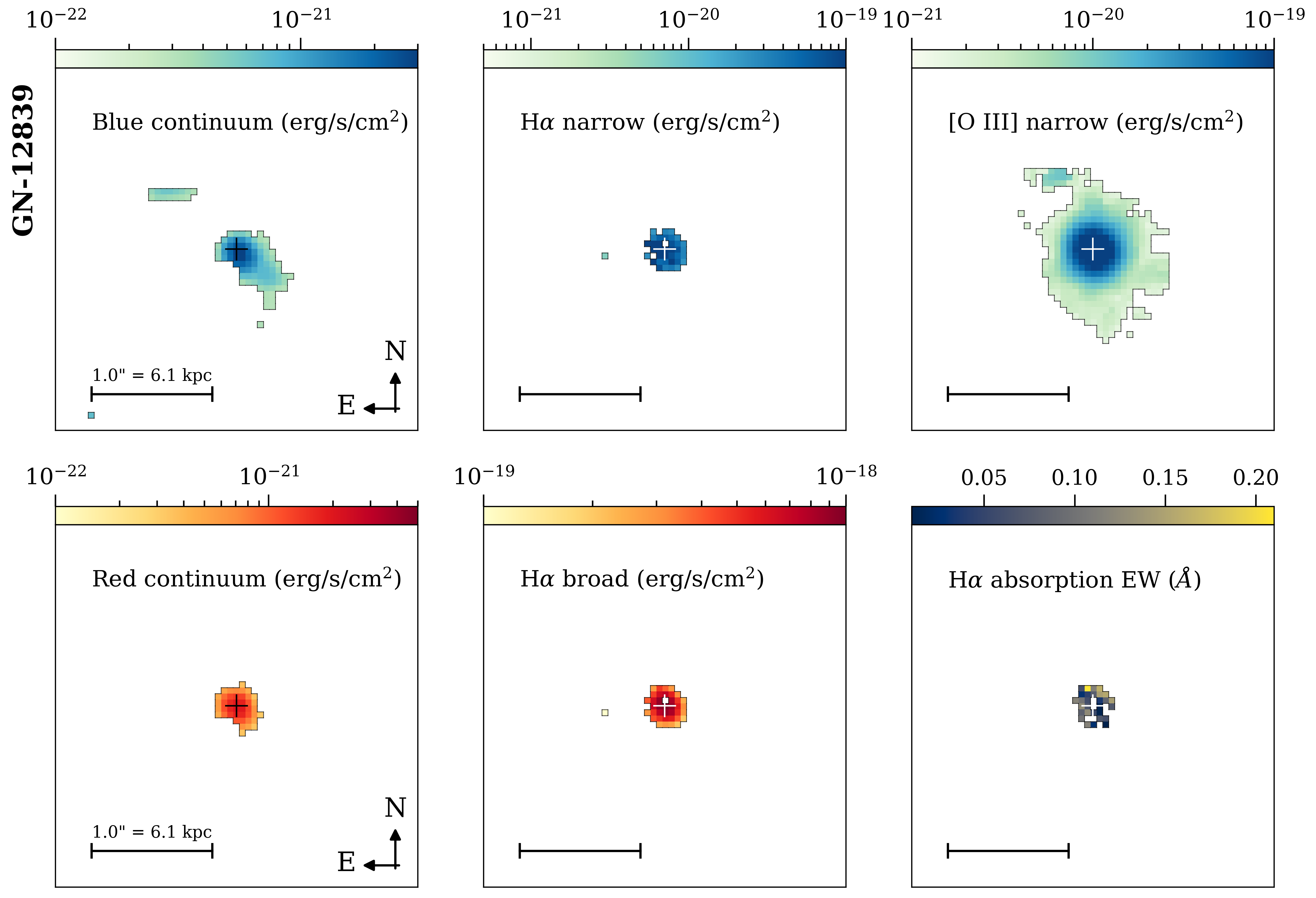}
    \end{center}
    \caption{Same as Figure~\ref{fig:gs13971}, but for \targGNB.}
    \label{fig:gn12839}
\end{figure*}

\begin{figure*}
    \begin{center} 
    \includegraphics[width=0.8\textwidth]{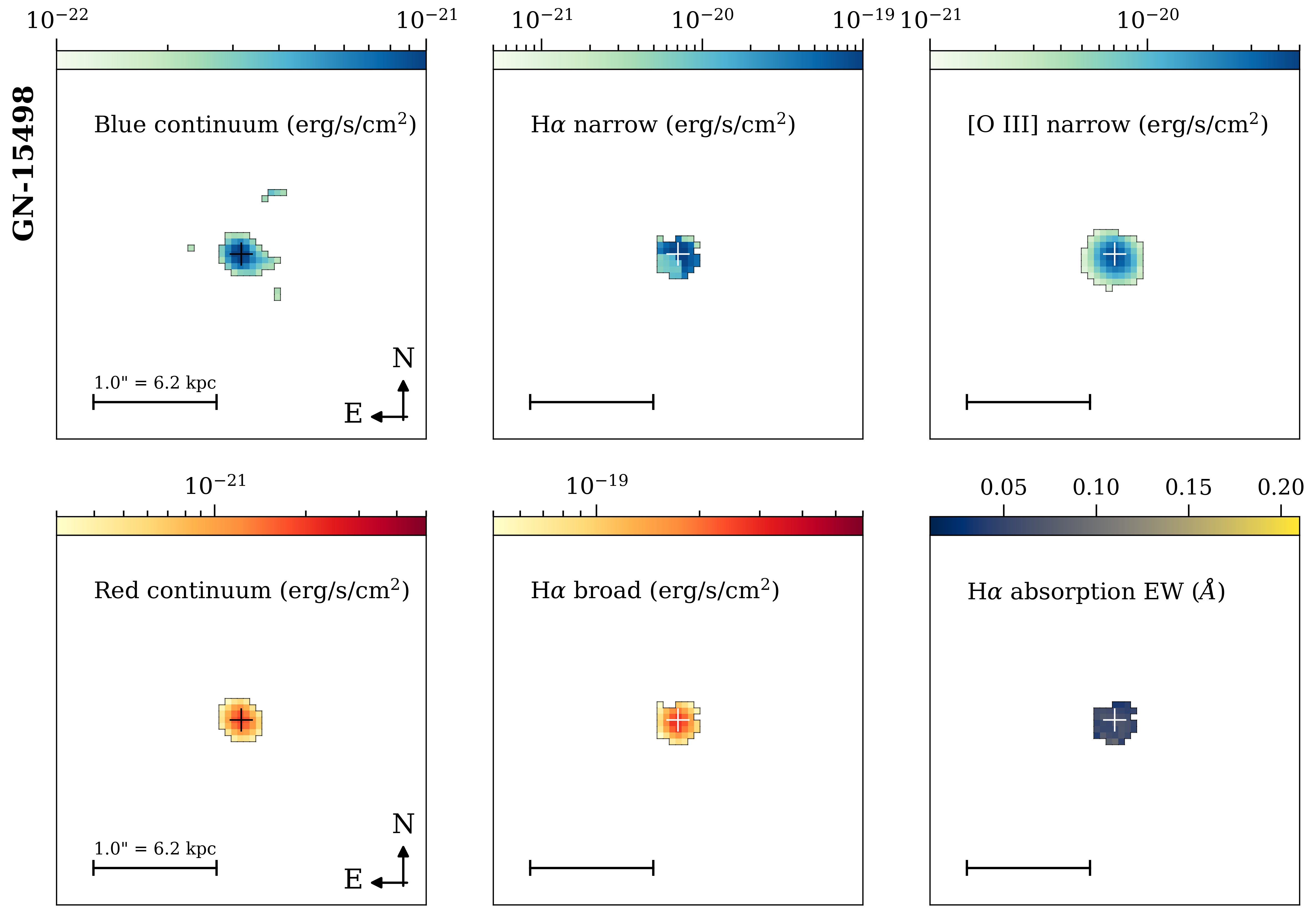}
    \end{center}
    \caption{Same as Figure~\ref{fig:gs13971}, but for \targGNC.}
    \label{fig:gn15498}
\end{figure*}

\begin{figure*}
    \begin{center} 
    \includegraphics[width=0.8\textwidth]{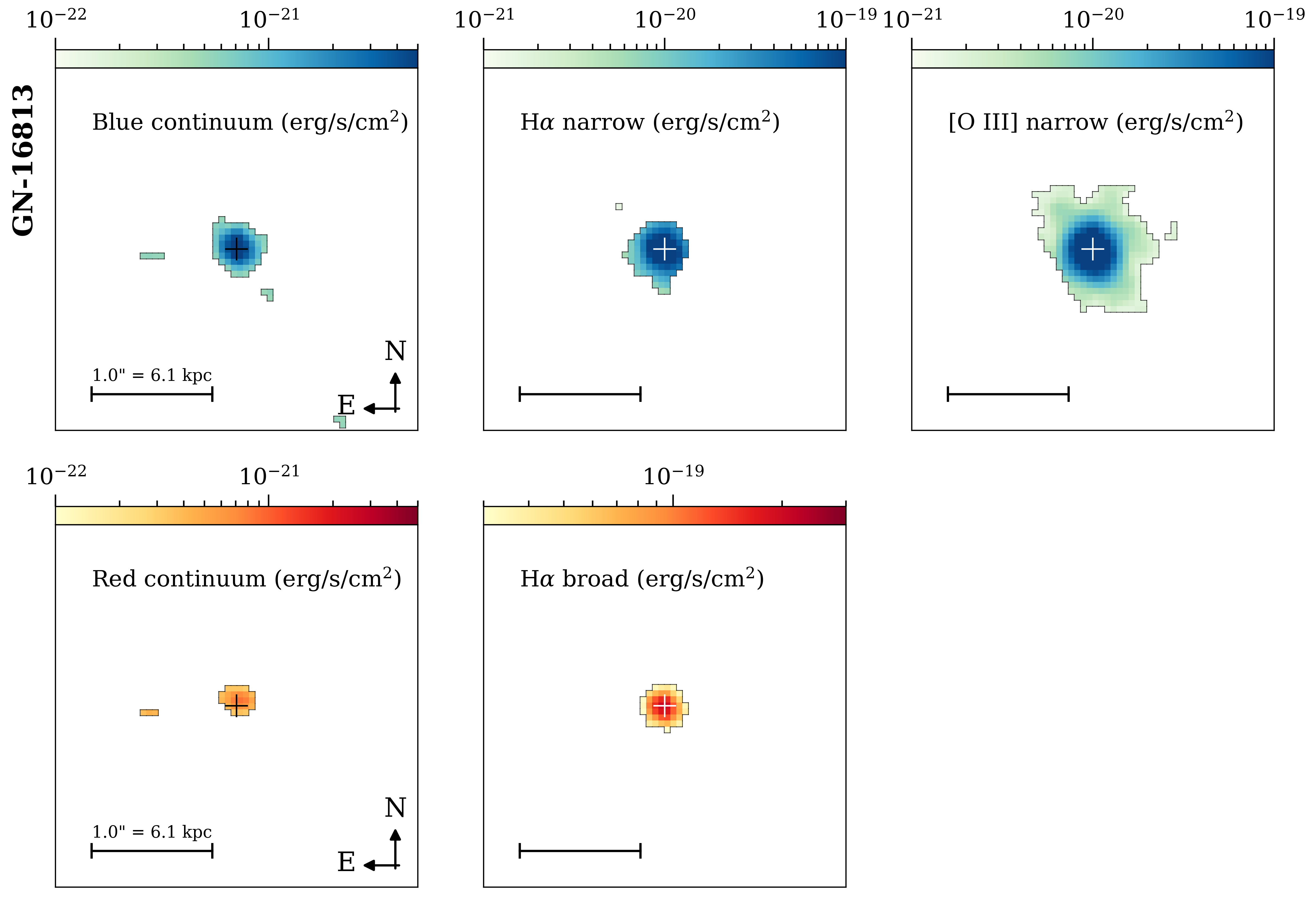}
    \end{center}
    \caption{Same as Figure~\ref{fig:gs13971}, but for \targGND.}
    \label{fig:gn16813}
\end{figure*}

\section{Could GS-13971 be in a merger?}\label{apdx:merger}
\begin{figure*}
    \begin{center} 
    \includegraphics[width=0.97\textwidth]{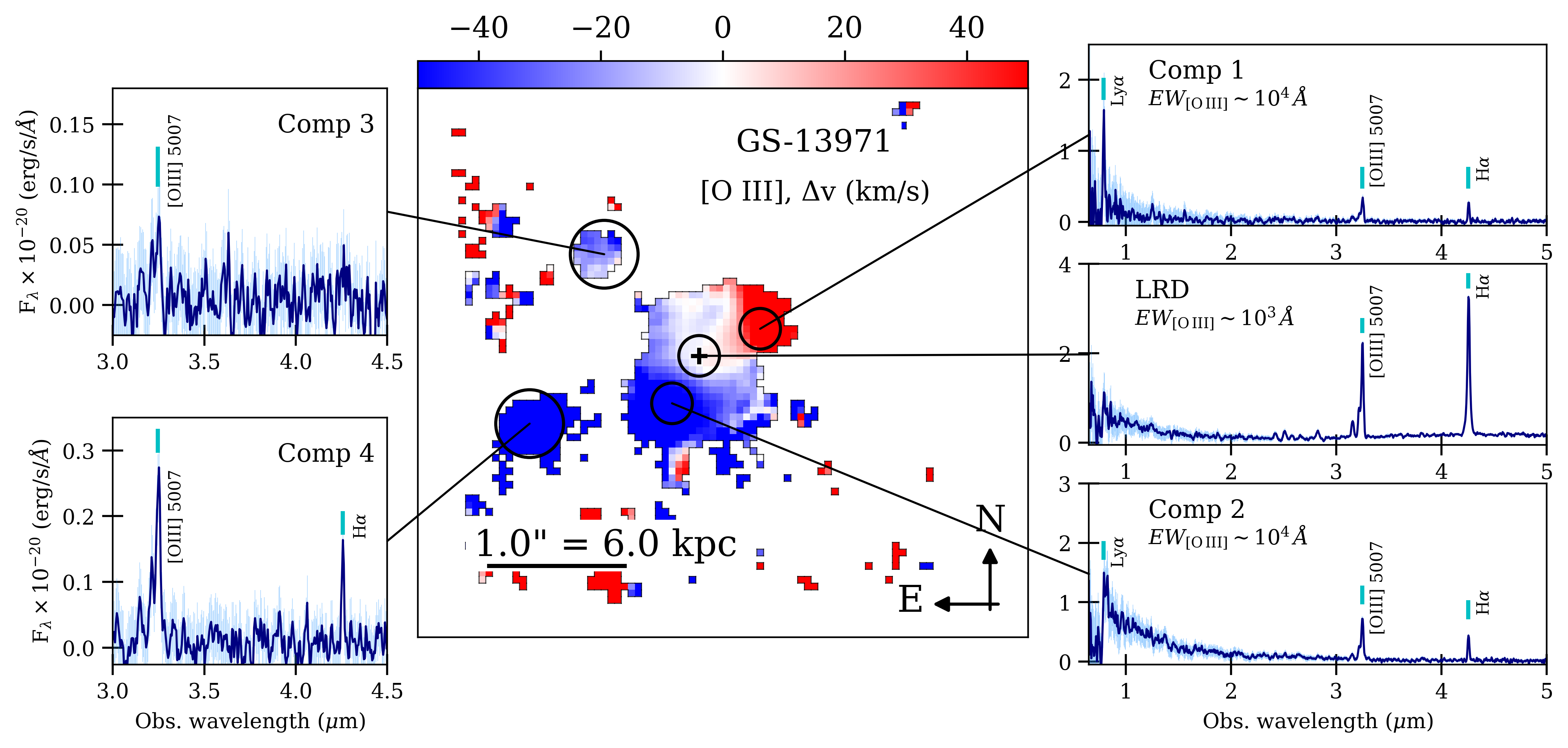}
    \end{center}
    \caption{We show the \OIII\ velocity map of \targGSA. We also show the observed aperture spectra, in linear scale, of the central LRD (marked by a black cross) and the four companions. The closest companions (Comp 1 and Comp 2) show distinct continuum and emission lines (e.g. \Lya, \OIII, and \Ha) from the central LRD. We also show the $\textrm{EW}_{\OIII}$ values corresponding to the central LRD, Comp 1, and Comp 2 apertures.}
    \label{fig:gsVelmap}
\end{figure*}

Here we investigate whether the extended ionized gas around \targGSA\ (Figure~\ref{fig:gs13971}) represents a large host galaxy or instead reflects merger activity involving the central LRD and nearby companions. By correlating the spatial distribution of \OIII\ and the blue continuum, we identify at least four narrow-line emitting companions surrounding \targGSA\ (labeled Comp 1-4 in Figure \ref{fig:gsVelmap}), as previously seen in NIRCam imaging \citep{Matthee2024}. 

Of particular interest is the ionized gas structure immediately surrounding the central LRD (Comp 1+2), projected within $0.65''$ ($\sim4$ kpc), which exhibits a northwest–southeast velocity gradient of approximately $\pm80~\rm km~s^{-1}$ (Figure~\ref{fig:gsVelmap}). We extract aperture spectra from the central LRD, Comp 1, and Comp 2 using a radius of $r=0.1''$ and find substantial differences in their  continuum and emission line properties. Only the LRD exhibits classic LRD-properties, including a v-shape continuum and broad emission lines, while the two companions only show a single-component power-law continua with prominent \Lya\ emission (Figure \ref{fig:gs13971}). Notably, Comp 2 has a brighter continuum than the central LRD. Although this analysis is not the main focus of this paper, we find that the observed velocity field is not well reproduced by a single rotating disk model. Taken together, these results suggest that the extended ionized gas around \targGSA\ is more consistent with a merger than a large rotating host galaxy. Interestingly, recent studies have suggested that LRDs may not host large reservoirs of cold gas \cite{XiaoM2025}, raising the possibility that \targGSA\ could represent a relatively gas-poor, dry merger. Further studies will be needed to better understand the nature of \targGSA.

\end{document}